\documentclass[conference]{IEEEtran}
\IEEEoverridecommandlockouts
\usepackage{cite}
\usepackage{amsmath,amssymb,amsfonts}
\usepackage{algorithmic}
\usepackage{graphicx}
\usepackage{textcomp}
\usepackage{xcolor}
\def\BibTeX{{\rm B\kern-.05em{\sc i\kern-.025em b}\kern-.08em
    T\kern-.1667em\lower.7ex\hbox{E}\kern-.125emX}}

\begin{document}

\title{Security of continuous-variable quantum key distribution against canonical attacks\\
\thanks{This work was funded by the European Union’s Horizon 2020 research
and innovation program under grant agreement No
820466 (CiViQ: “Continuous Variable Quantum Communications”).}
}

\author{\IEEEauthorblockN{Panagiotis Papanastasiou}
\IEEEauthorblockA{\textit{Computer Science} \\
\textit{University of York}\\
York, United Kingdom \\
panagiotis.papanastasiou@york.ac.uk}
\and
\IEEEauthorblockN{Carlo Ottaviani}
\IEEEauthorblockA{\textit{Computer Science} \\
\textit{University of York}\\
York, United Kingdom  \\
carlo.ottaviani@york.ac.uk}
\and
\IEEEauthorblockN{ Stefano Pirandola}
\IEEEauthorblockA{\textit{Computer Science} \\
\textit{University of York}\\
York, United Kingdom  \\
stefano.pirandola@york.ac.uk}}

\maketitle

\begin{abstract}
We investigate the performance of Gaussian-modulated coherent-state QKD protocols in the presence of canonical attacks, which are collective Gaussian attacks resulting in Gaussian channels described by one of the possible canonical forms. We present asymptotic key rates and then we extend the results to the finite-size regime using a recently-developed toolbox for composable security. 
\end{abstract}

\begin{IEEEkeywords}
Continuous variables, quantum key distribution, Gaussian modulation, finite-size effects, composable security
\end{IEEEkeywords}

\section{Introduction}
A quantum key distribution (QKD) protocol describes the communication steps performed by two remote authenticated parties to establish a shared key even though the link between them is potentially compromised~\cite{revQKD}. The information-theoretic security of such a protocol is granted by the  laws of nature (quantum mechanics)~\cite{noclone0,noclone}. The first protocols designed were based on discrete variable (DV) systems, while more recently proposed  protocols use continuous variables (CV), i.e., the position and momentum quadratures of the bosonic modes of the electromagnetic field~\cite{Braunstein_rev,Stefano_rev}. In particular, CV-QKD protocols using Gaussian modulation of coherent states for the encoding of information~\cite{GG02,RR_protocol,weedbrook2004noswitching} can be easily implemented using the current telecommunication infrastructure and may achieve high rates close to the PLOB  bound for repeaterless quantum communications in a lossy channel~\cite{PLOB}. More specifically, these protocols can be considered as coming from a single scheme with different aspects~\cite{Stefano_rev}: reverse reconciliation (RR) or direct reconciliation (DR), with homodyne or heterodyne decoding measurement. 

Their security analysis was first studied for asymptotic key rates under the assumption of collective Gaussian attacks~\cite{opt_Gaussian_attack1,opt_Gaussian_attack2}, completely characterized by Ref.~\cite{Stefano_CF}. Later, security was extended to the finite size regime~\cite{Antony_cpe, UsenkoFNSZ, finite-size thermal} and to a general composable framework~\cite{Leverier_definetti,free space}, including free-space~\cite{free space} and satellite-based scenarios~\cite{SatQKD}. Proof-of-principle and in-field experiments have been recently demonstrated in long ground-based fiber connections~\cite{HuangExp,ZhangExpI,ZhangExpII}. As pointed out in Ref.~\cite{Stefano_CF}, single-mode Gaussian channels and the corresponding collective Gaussian attacks can be classified in different canonical forms. One of these forms is represented by the thermal-loss (attenuation) channel and the associated collective entangling-cloner attack, typically assumed in CV-QKD security proofs.

Here we present the asymptotic secret key rates of the Gaussian-modulated coherent-state protocols with respect to the other canonical forms. Besides the attenuation channel, these include the amplifying channel, the additive classical-noise channel and other more exotic Gaussian channels~\cite{Stefano_rev,Stefano_CF}. Then, using the toolbox of Ref.~\cite{free space} for composable security under general channel conditions, we extend the analysis of the amplifying and classical-noise channels to include finite-size effects and composable security. 

After a short description of the canonical forms in Sec.~\ref{sec:Canforms}, in Sec.~\ref{sec:Protocol} we describe the security analysis in the asymptotic regime in the presence of a generic canonical form for the cases of for homodyne or heterodyne   protocol in RR/DR. In Sec.~\ref{sec:rates}, we present the results of the previous analysis specified for each canonical form by assuming ideal reconciliation efficiency and large modulation. In Sec.~\ref{sec:PE}, we perform the parameter estimation (PE) following Refs.~\cite{UsenkoFNSZ,finite-size thermal} and in Sec.~\ref{sec:composable} we compute the composable key rates using Ref.~\cite{free space}.

\section{Canonical Forms\label{sec:Canforms}}
Recall that a Gaussian channel $\mathcal{G}(\mathbf{T},\mathbf{N},\mathbf{d})$ acting on a single mode, for $\mathbf{T},\mathbf{N}$ $2\times 2$ real matrices and $\mathbf{d}$ an $\mathbb{R}^2$ vector, is a completely positive trace-preserving map that maintains the Gaussian statistics of the input state. It can be mapped to its canonical form $\mathcal{C}$, which is a Gaussian channel with $\mathbf{d}=0$ and $\mathbf{T}_c$, $\mathbf{N}_c$ diagonal, by $\mathcal{G}=\mathcal{U}_A \circ \mathcal{C} \circ \mathcal{U}_B$, where $\mathcal{U}_A$ and $\mathcal{U}_B$ are Gaussian unitaries. One can reduce the description of $\mathbf{T}_c$, $\mathbf{N}_c$ to three symplectic invariants: the generalized transmission $\tau=\text{det}\mathbf{T}$, for $-\infty<\tau<\infty$, the rank $r=(\text{rk}(\mathbf{T})\text{rk}(\mathbf{N}))/2$ for $r=0,1,2$ and the temperature $\bar{n}$, connected to $\text{det}\mathbf{N}$. 

According to the first two parameters, the canonical forms can be grouped into different classes:
The class $A_1$ for $\tau=0$, $r=0$, which replaces the input states with thermal states (completely depolarizing channel). The classes $A_2$ and $B_1$ for $\tau=0$, $r=1$ and $\tau=1$, $r=1$ transforming the quadratures asymetrically.  $B_2$ is the additive classical-noise channel for $\tau=1$ and $r=2$ and it collapses to the identity channel for $\bar{n}=0$.
Class $C$ is connected to channels with transmissivity, i.e., $0<\tau\neq 1$ and $r=2$, with the subcases $\tau<1$ (attenuation channel) and $\tau>1$ (amplifying channel). Finally, the class $D$, where its output can be seen as complementary to the amplifying channel and is connected to negative transmissivities.

Via the Stinespring dilation, one can represent the canonical form $\mathcal{C}(\tau,r,\bar{n})$ with a unitary symplectic transformation $\mathcal{L}(\tau,r)$ mixing the input state and a two-mode squeezed-vacuum (TMSV) state with variance $\omega=2\bar{n}+1$, which describes the environment. In more detail, apart from the class $B_2$, we have that  $\mathcal{L}(\tau,r)=\mathcal{M}(\tau,r)\oplus \mathbf{I}$ where $\mathcal{M}(\tau,r)$ is a symplectic form interacting only with the input state and one mode from the TMSV state, with the other mode being subject to the identity $\mathbf{I}$. For the class $B_2$, we adopt a description using the attenuation channel as we will see later.

The unitary dilation of the canonical form represents the Gaussian interaction performed by the eavesdropper that controls the TMSV state of the environment~\cite{Stefano_CF}. After interaction with the input mode, the environmental output is stored in a quantum memory, that will be subject to a joint and optimal measurement (collective attack).


\section{Aspects of the protocol scheme\label{sec:Protocol}}
Alice picks randomly $2N$ samples $\{x_i\}$ from the variable $x$ distributed according to the normal distribution 
\begin{equation}
p(x)=(\sqrt{2\pi V_A})^{-1}\exp\left[-x^2/(2V_A)\right]
\end{equation}
with zero mean and variance $V_A$. Then she
modulates mode $A$ carrying coherent states $|\alpha\rangle$ according to these samples with
\begin{equation}
\alpha=(q_A+\mathrm{i}p_A)/2=(x_{2j-1}+\mathrm{i}x_{2j})/2,
\end{equation}
where $q_A$ and $p_A$ are the encoding on the quadratures and $j=1,\dots,N$.
In the asymptotic regime ($N \gg 1$), the covariance matrix (CM) of Alice's ensemble state is given by $\mathbf{V}_A= \mu \mathbf{I}$ with $\mathbf{I}= \text{diag} \{1,1\}$ and $\mu=V_A+1$.
Mode $A$ is traveling through a quantum channel modeled by one of the canonical forms~\cite{Stefano_rev,Stefano_CF}. In particular, Eve's system is described by two modes $E$ and $e$ in a TMSV state  with variance $\omega$ and covariance matrix
\begin{equation}
\mathbf{V}_{Ee}=\begin{pmatrix}
\omega \mathbf{I}&\sqrt{\omega^2-1} \mathbf{Z}\\
\sqrt{\omega^2-1}\mathbf{Z}&\omega \mathbf{I}
\end{pmatrix},
\end{equation}
where  $\mathbf{Z}=\text{diag} \{1,-1\}$. Mode $E$ is mixed with $A$ via a canonical form whose dilation is represented by a symplectic matrix $\mathcal{M}$ (e.g., this is a beam-splitter transformation in the case of an attenuation channel). One output mode $B$ goes to Bob, while the other $E'$ goes to Eve. Eve's idler mode $e$ and mode $E'$ are kept in a quantum memory for a later optimal measurement. Then the CM for modes $B$, $E'$, and $e$ is given by
\begin{equation}
\mathbf{V}_{BE'e}=(\mathcal{M}^\mathsf{T} \oplus \mathbf{I}) \left(\mathbf{V}_A\oplus\mathbf{V}_{Ee}\right) (\mathcal{M} \oplus \mathbf{I}),
\end{equation}
which can be expressed as follows
\begin{equation}\label{eq:CMBE'e}
\mathbf{V}_{BE'e}=\begin{pmatrix}
\mathbf{V}_B  & \mathbf{C}_{BE'}&\mathbf{C}_{Be}\\
\mathbf{C}_{BE'}&\mathbf{V}_{E'}&\mathbf{C}_{E'e}\\
\mathbf{C}_{Be}&\mathbf{C}_{E'e}&\mathbf{V}_{e}
\end{pmatrix}.
\end{equation}

From the previous CM, we can derive the CM of Eve's average state by tracing out mode $B$ and Bob's CM by tracing out $E'e$ respectively. Therefore, we obtain
\begin{equation}\label{eq:average CM}
\mathbf{V}_{E'e}=\begin{pmatrix}
\mathbf{V}_{E'} &\mathbf{C}_{E'e} \\
\mathbf{C}_{E'e}&\mathbf{V}_{e}  
\end{pmatrix},~~
\mathbf{V}_B=\text{diag}\{V^q_B(V_A),V^p_B(V_A)\}
\end{equation}
where $\mathbf{V}_{E'}=\text{diag}\{V^q_{E'}(V_A),V^p_{E'}(V_A)\}$ is a function of $V_A$. In general, the canonical forms 
may treat the quadratures $q$ and $p$ asymmetrically resulting in different variances $V_B^q$ and $V_B^p$ or $V_E^q$ and $V_E^p$
respectively. Note that for the class $C$ and the classical-noise channel the treatment is symmetric so we have $V_B^q=V_B^p=V_B$ and $V_{E'}^q=V_{E'}^p=V_{E'}$.

In the homodyne protocol, Bob measures either the $q$-quadrature or $p$-quadrature of the arriving mode with outcome $q_B$ or $p_B$ respectively. He informs Alice about the choice of quadrature and then she keeps only the relevant encoding $q_A$ or $p_A$ respectively (shifting the outcomes). In contrast, in the heterodyne protocol, Bob measures both quadratures and Alice's encoding is described by the pair $q_A,p_A$ and Bob's outcome by $q_B,p_B$.

For the homodyne protocol in DR, we derive Eve's conditional CM $\mathbf{V}_{E'e|q_A}$ (respectively $\mathbf{V}_{E'e|p_A}$) on Alice's encoding $q_A$ (or $p_A$) given by~(\ref{eq:average CM}) up to the replacement of $\mathbf{V}_{E'}$ with $\text{diag}\{V^q_{E'}(0),V^p_{E'}(V_A)\}$ (respectively with $\text{diag}\{V^q_{E'}(V_A),V^p_{E'}(0)\}$); for the heterodyne protocol, the conditional CM  $\mathbf{V}_{E'e|q_A,p_A}$ is given by replacing $\mathbf{V}_{E'}$ by $\text{diag}\{V^q_{E'}(0),V^p_{E'}(0)\}$ in the same equation.

Let us now compute Eve's conditional CM on Bob's measurement outcome $l_B$, with $l$ equal to either $q$ or $p$ for a different quadrature, in RR. For the homodyne protocol, we obtain~\cite{Stefano_rev}
 \begin{equation}
\mathbf{V}_{E'e|l_B}=\mathbf{V}_{E'e}-\mathbf{C}_{BE'e}^\mathsf{T}\left(\Pi_{l}\mathbf{V}_B\Pi_{l}\right)^{-1}\mathbf{C}_{BE'e},
\end{equation}
where $\mathbf{C}_{BE'e}=\begin{pmatrix} \mathbf{C}_{BE'} \\ \mathbf{C}_{Be} \end{pmatrix}$, $\Pi_q=\text{diag}\{1,0\}$, $\Pi_p=\text{diag}\{0,1\}$, and $(.)^{-1}$ corresponds here to the calculation of the pseudo-inverse. If Bob's measurement is a heterodyne measurement then the conditional CM is given by~\cite{Stefano_rev}
\begin{equation}
\mathbf{V}_{E'e|q_B,p_B}=\mathbf{V}_{E'e}-\mathbf{C}_{BE'e}^\mathsf{T}\left(\mathbf{V}_B +\mathbf{I}\right)^{-1}\mathbf{C}_{BE'e}.
\end{equation}

Let us assume now a very large number of exchanged signals ($N\gg1$). Then the mutual information between the encoding $q_A$ or $p_A$ and the outcome $q_B$ or $p_B$ for the homodyne protocol is given by
\begin{equation}
I(\mu,\tau,\omega)=\frac{1}{2}\left(\frac{1}{2}\log_2\frac{V^q_{B}}{V^q_{B|q_A}}+\frac{1}{2}\log_2\frac{V^p_{B}}{V^p_{B|p_A}}\right),
\end{equation}
for $V^l_{B|l_A}=V^l_{B}(0)$, where we have assumed that half of the times Bob's outcome is $q_B$ and otherwise $p_B$. On the other hand, the mutual information between the encoding $q_A,p_A$  and the outcome $q_B, p_B$  for the heterodyne protocol is given by
\begin{equation}
I(\mu,\tau,\omega)=\frac{1}{2}\left(\log_2\frac{V^q_{B}+1}{V^q_{B|q_A}+1}+\log_2\frac{V^p_{B}+1}{V^p_{B|p_A}+1}\right).
\end{equation}


Eve's Holevo information is calculated by the symplectic spectrum $\boldsymbol{\nu}_{E'e}$ of the CM $\mathbf{V}_{E'e}$ and the spectra, $\boldsymbol{\nu}_{E'e|q_A}$, $\boldsymbol{\nu}_{E'e|p_A}$ and $\boldsymbol{\nu}_{E'e|q_A,p_A}$ or $\boldsymbol{\nu}_{E'e|q_B}$, $\boldsymbol{\nu}_{E'e|p_B}$ and $\boldsymbol{\nu}_{E'e|q_B, p_B}$, associated with the conditional CMs in DR or RR respectively. More specifically, for the homodyne protocol, we have that
\begin{align}
\chi(\mu,\tau,\omega)&=\sum_{i=1,2} h\left([\boldsymbol{\nu}_{E'e}]_i\right)\notag\\
&-\frac{1}{2}\left(\sum_{i=1,2} h\left([\boldsymbol{\nu}_{E'e|q_\gamma}]_i\right)+\sum_{i=1,2} h\left([\boldsymbol{\nu}_{E'e|p_\gamma}]_i\right)\right),
\end{align}
while for the heterodyne protocol
\begin{align}
\chi(\mu,\tau,\omega)&=\sum_{i=1,2} h\left([\boldsymbol{\nu}_{E'e}]_i\right)\notag
-\sum_{i=1,2} h\left([\boldsymbol{\nu}_{E'e|q_\gamma,p_\gamma}]_i\right),
\end{align}
where
\begin{equation}
h(x)=\frac{x+1}{2}\log_2\frac{x+1}{2}-\frac{x-1}{2}\log_2\frac{x-1}{2},
\end{equation}
with $\gamma$ being either $A$ or $B$ for the protocol in DR or RR respectively,
Then the asymptotic secret key rate is obtained by~\cite{revQKD}
\begin{equation}\label{eq:asym_rate}
R(\mu,\tau,\omega)=\zeta I(\mu,\tau,\omega)-\chi(\mu,\tau,\omega),
\end{equation}
where $\zeta$ is the reconciliation efficiency parameter.

\section{\label{sec:rates}Asymptotic key rates}

Here we calculate the asymptotic secret key rate for each of the canonical forms assuming an ideal reconciliation efficiency $\zeta=1$ and the large modulation limit $\mu \rightarrow \infty$. In fact, we present results in detail for the practical cases of attenuation, amplifying, and classical-noise channel. Class $B_1$ is always secure (see Appendix for more details) while classes $D$ and $A_2$ do not provide a secret key rate, i.e., for any set of parameters describing the corresponding canonical form the parties cannot extract a secret key. The whole class of such channels have the property of anti-degradability~\cite{Stefano_rev}: In terms of cryptography, the eavesdropper (Eve) can obtain the receiver's (Bob's) state by applying a CPT map on the state of the environment forbidding the secret key extraction. However, for classes with members that may hold this property or not, e.g., the attenuation channel for $\tau<1/2$ against the cases with $\tau\geq1/2$, the RR can provide a remedy. 
\subsection{C class}
\begin{figure}[t]
\centering
\includegraphics[width=0.4\textwidth]{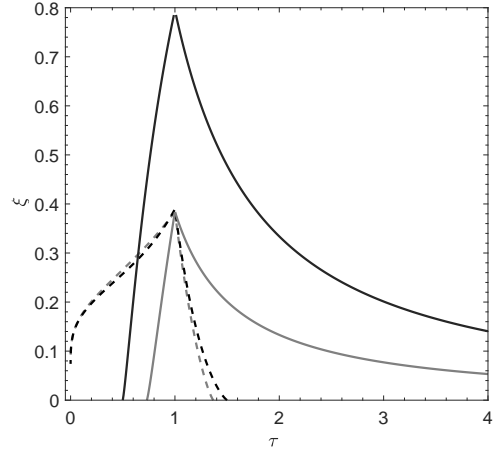}
\caption{\label{fig:thresholds} The asymptotic security thresholds of the C class for transmisivities $\tau>0$ ($\tau\neq 1$) with respect to the excess noise $\xi$, where the reconciliation has been considered ideal and $\mu\rightarrow \infty$. We plot the homodyne protocol in DR (black solid line) and in  RR (black dashed line) and the heterodyne protocol in DR (gray solid line) and in RR  (gray dashed line). The instances with high excess noise (above the threshold lines) give no secret key rate.}
\end{figure}

\begin{figure}[t]
\centering
\includegraphics[width=0.4\textwidth]{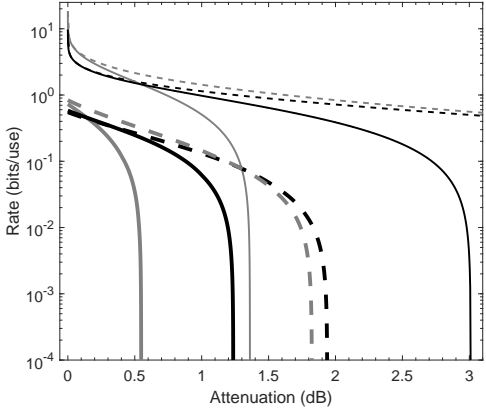}
\caption{\label{fig:Att_asy}Secret key rate versus attenuation in dB for an attenuation channel. With thin lines, we plot the asymptotic rate for the homodyne  protocol in DR (black solid) and in RR (black dashed) and for the heterodyne protocol in DR (gray solid) and in RR (gray dashed) for $\xi=0$, $\zeta=1$, and $\mu \rightarrow \infty$. For the composable secret key rates (corresponding thick lines), we have assumed channel excess noise $\xi=0.01$ and  conservative values for the parameters $\zeta=0.9$, $p_\text{EC}=0.8$, and $N=10^6$. We have optimized over the ratio $r$ and the modulation $V_A$ with $\epsilon_\text{PE}\approx 10^{-10}$, $\epsilon_\text{s}=\epsilon_\text{h}=10^{-20}$, and $d=2^5$. The plots of the asymptotic key rate evaluate the security of the protocol and the associated performance taking into account only theoretical aspects, e.g.  the kind of attack, focusing more on the quantum communication part of the protocol. On the contrary, the finite-size analysis in a composable framework takes also into account the classical post-processing parts of the protocol providing with a performance close to a practical implementation usually expected to be worse than the ideal case as it is supported by the plots in thick lines compared with the corresponding cases in thin lines. We observe here that the heterodyne protocols behave better in closer distances (higher signal to noise ratio) compared with the homodyne protocols since they can take into advantage the double encoding into the same signal. Despite this fact, the homodyne protocols have achievable rates in longer distances. In fact, they behave better  against the excess noise in long distances and, in particular the RR protocol, against the parameter estimation effects connected to the excess noise and transmissivity.}
\end{figure}

\begin{figure}[t]
\centering
\includegraphics[width=0.4\textwidth]{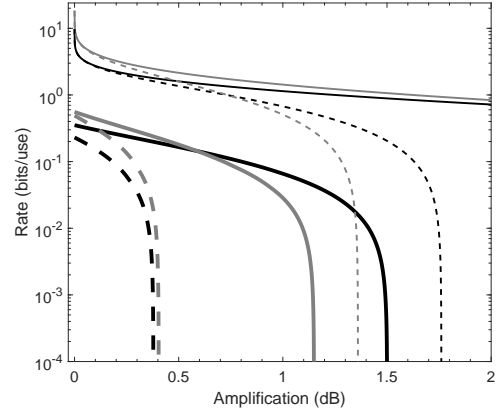}
\caption{\label{fig:Amp_asy} Asymptotic key rates in the presence of an amplifying channel. With thin lines, we plot the asymptotic rate for the homodyne  protocol in DR (black solid) and in RR (black dashed) and for the heterodyne protocol in DR (gray solid) and in RR (gray dashed) for $\xi=0$, $\zeta=1$ and $\mu \rightarrow \infty$. For the composable secret key rates (corresponding thick lines), we have assumed channel excess noise $\xi=0.01$ and  conservative values for the parameters $\zeta=0.9$, $p_\text{EC}=0.8$, and $N=10^6$. We have optimized over the ratio $r$ and the modulation $V_A$ with $\epsilon_\text{PE}\approx 10^{-10}$, $\epsilon_\text{s}=\epsilon_\text{h}=10^{-20}$, and $d=2^5$. Here we observe that the RR and DR protocols have the opposite behaviour compared with the attenuation channel case, i.e., we have smaller achievable rate distances for the RR protocols instead for the DR protocols.  Comparing also the composable rates of the RR protocols, it seems that in the regime of $N=10^6$, the homodyne protocol cannot surpass the performance of the heterodyne protocol due to the fact that the gain variance plays an important role in the amplifying channel: the coefficient in front of the gain variance in~(\ref{eq:sigmaXihom_amp}) is double of its counterpart in~(\ref{eq:sigmaXihet_amp}).}
\end{figure}

The symplectic matrix associated with the dilation of the  C class is 
\begin{equation}
\mathcal{M}_\text{Att}(0<\tau<1)=\begin{pmatrix}
\sqrt{\tau}\mathbf{I} & \sqrt{1-\tau}\mathbf{I}\\
-\sqrt{1-\tau}\mathbf{I} & \sqrt{\tau}\mathbf{I}
\end{pmatrix}
\end{equation}
and 
\begin{equation}
\mathcal{M}_\text{Amp}(\tau>1)=\begin{pmatrix}
\sqrt{\tau}\mathbf{I} & \sqrt{\tau-1}\mathbf{Z}\\
\sqrt{\tau-1}\mathbf{Z} & \sqrt{\tau}\mathbf{I}
\end{pmatrix}
\end{equation}
for the attenuation and amplifying channel respectively.
Following the steps in Sec.~\ref{sec:Protocol}, one easily obtains the secret key rates for the homodyne (hom) and heterodyne (het) protocols
in DR ($\blacktriangleright$) and RR ($\blacktriangleleft$). We have
\begin{align}
R_\text{hom}^\blacktriangleright (\tau,\omega)&=\frac{1}{2}\log_2\frac{\tau\left(\tau\omega+|1-\tau|\right)}{|1-\tau|(\tau+|1-\tau|\omega)}\notag\\&~~~~-h(\omega)+h\left(\sqrt{\frac{\omega(\tau+|1-\tau|\omega)}{|1-\tau|+\tau\omega}}\right),\label{eq:AttDRhom}\\
R_\text{hom}^\blacktriangleleft(\tau,\omega)&=\frac{1}{2}\log_2\frac{\omega}{|1-\tau|(\tau+(|1-\tau|)\omega)}-h(\omega),\label{eq:AttRRhom}\\
R_\text{het}^\blacktriangleright (\tau,\omega)&=\log_2\frac{2\tau}{\mathrm{e}|1-\tau|\left(\tau+|1-\tau|\omega+1\right)}\notag\\&~~~~-h(\omega)+h\left(\tau+|1-\tau|\omega\right),\label{eq:AttDRhet}\\
R_\text{het}^\blacktriangleleft(\tau,\omega)&=\log_2\frac{2\tau}{\mathrm{e}|1-\tau|(\tau+|1-\tau|\omega+1)}\notag\\&~~~~-h(\omega)+h\left(\frac{1+|1-\tau|\omega}{\tau}\right)\label{eq:AttRRhet}.
\end{align}
 In Fig.~\ref{fig:thresholds}, we plot the security threshold for each of the cases above with respect to transmissivity and excess noise $\xi=\frac{|1-\tau| (\omega-1)}{\tau}$.
Then, in Fig.~\ref{fig:Att_asy}, for  $\omega:=1$ (no thermal noise), $\zeta=1$, and $\tau:=10^{\frac{-L}{10}}$ where $L$ is the attenuation in dB, we plot~(\ref{eq:AttDRhom}),~(\ref{eq:AttDRhet}),~(\ref{eq:AttRRhom}) and~(\ref{eq:AttRRhet}). In Fig.~\ref{fig:Amp_asy}, we plot the same cases for $\tau:=10^{\frac{L}{10}}$ with $L$ being the gain in dB.

\subsection{Classical-noise channel}
\begin{figure}[t]
\centering
\includegraphics[width=0.4\textwidth]{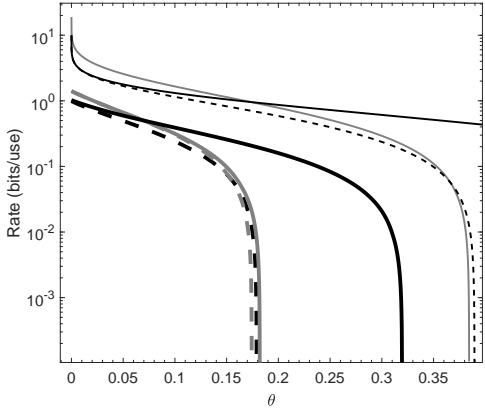}
\caption{\label{fig:Clas_hom}Secret key rate for a classical-noise channel against the classical thermal noise $\theta$. With thin lines, we plot the asymptotic rate for the homodyne  protocol in DR (black solid) and in RR (black dashed) and for the heterodyne protocol in DR  (gray solid) and in RR (gray dashed) for $\xi=0$, $\zeta=1$ and $\mu \rightarrow \infty$. Note that the lines for the heterodyne protocol in DR and RR coincide. For the composable secret key rates (corresponding thick lines), we have assumed channel excess noise $\xi=0.01$ and  conservative values for the parameters $\zeta=0.9$, $p_\text{EC}=0.8$, and $N=10^6$. We have optimized over the ratio $r$ and the modulation $V_A$ with $\epsilon_\text{PE}\approx 10^{-10}$, $\epsilon_\text{s}=\epsilon_\text{h}=10^{-20}$, and $d=2^5$. Here we observe that the most robust protocol against classical thermal noise is the homodyne protocol in DR while the other cases have similar performance.}
\end{figure}

To simulate a Gaussian channel with additive classical-noise, we adopt the symplectic matrix of a beam splitter 
\begin{equation}
\mathcal{M}_\text{Att}(0<\tau<1)=\begin{pmatrix}
\sqrt{\tau}\mathbf{I} & \sqrt{1-\tau}\mathbf{I}\\
-\sqrt{1-\tau}\mathbf{I} & \sqrt{\tau}\mathbf{I}
\end{pmatrix}
\end{equation}
and take the joint limits for $\tau \rightarrow 1$ and $\omega \rightarrow \infty$ so that $(1-\tau)\omega=\theta$, for some constant variance $\theta$ of the additive noise.
The corresponding secret key rates are given by
\begin{align}
&R_{\text{hom}}^\blacktriangleright(\theta)=\log_2\left(\frac{2}{e \sqrt{\theta(\theta+1)}}\right)+h(\sqrt{1+\theta}),\label{eq:ClasDRhom}\\
&R_{\text{hom}}^\blacktriangleleft(\theta)=\log_2 \left(\frac{2}{e \sqrt{\theta  (\theta +1)}}\right),\label{eq:ClasRRhom}\\
&R_{\text{het}}^\blacktriangleright(\theta)=R_{\text{het}}^\blacktriangleleft(\theta)=\log_2 \left(\frac{4}{e^2 \theta  (\theta +2)}\right)+h(\theta+1),\label{eq:ClasDRhet}
\end{align} 
We plot~\eqref{eq:ClasDRhom},~\eqref{eq:ClasRRhom} and~\eqref{eq:ClasDRhet} in Fig.~\ref{fig:Clas_hom}.

\section{\label{sec:PE}Parameter estimation}
\subsection{\label{sec:PEAtt}Attenuation channel}
Let us assume a protocol with homodyne detection. Here Bob's measurement outcome is described by the generic variable
\begin{align}
y=&\sqrt{\tau} x+z
\end{align}
where $y$ describes either the outcome connected with the quadrature $q$ or $p$. Accordingly, the variable $x$ describes Alice's encoding while 
\begin{align}
z~=\sqrt{\tau}x_s+\sqrt{1-\tau} x_o +x_\Xi,
\end{align}
is a variable representing the noise detected by Bob.  The variables $x_s$ and $x_o$ have equal variance $V_s=V_o=1$ describing the quantum shot noise, and the variable $x_\Xi$ with variance $\Xi:=\tau\xi$ describes the excess noise of the channel $\xi=\frac{(1-\tau)(\omega-1)}{\tau}$. 
Therefore we obtain the noise variance
\begin{align}\label{eq:signalnoiseAtt}
\sigma^2_z=\Xi+1.
\end{align}



Based on the previous analysis and assuming $m$ signals for PE, we derive the variances of the maximum likelihood estimators (MLEs) $\widehat{\tau}$ and $\widehat{\Xi}$ of the transmissivity and excess noise according to Ref.~\cite{finite-size thermal}. Therefore,  the worst case scenario values for the channel parameters are given by  
\begin{align}
\tau_m=\tau-w \sigma_\tau,~~\Xi_m=\Xi+w \sigma_\Xi,
\end{align}
with  $w=\sqrt{2}\text{erf}^{-1}(1-\epsilon_{\mathrm{PE}})$, as the extremal values of the intervals defined by the estimator variances 
\begin{align}
\sigma_\tau^2=\frac{4\tau^2}{m}\left( 2+\frac{\sigma_z^2}{\tau V_A}\right),~~\sigma_\Xi^2=2 \frac{\sigma_z^4}{m}\label{eq:vartauh1},
\end{align}
where $\text{erf}(.)$ is the error function and $\epsilon_{\mathrm{PE}}$ is the associated error probability.

In the heterodyne protocol, Bob mixes the incoming mode $B$ with a vacuum mode in a balanced beam splitter. Then he  applies two conjugate homodyne detections to
the beam-splitter outputs. Due to the presence of the extra vacuum mode, the outputs have an increased noise variance by $1$ shot noise units compared with the protocol using homodyne measurement. In addition, there is an estimator for $\tau$ and $\Xi$ from each one of the quadratures. These are optimally combined and give the variances

\begin{equation}
\sigma_\tau^2=\frac{2\tau^2}{m}\left( 2+\frac{\sigma^2_z+1}{\tau V_A}\right)~\text{and}~\sigma^2_\Xi=\frac{(\sigma^2_z+1)^2}{m}.
 \end{equation}
Finally, the key rate in Eq.~(\ref{eq:asym_rate}) is expressed via the parameter $\Xi$ as $\tilde{R}(\mu,\tau,\Xi)=R(\mu,\tau,\omega)$ and by setting the worst case scenario values one obtains the secret key rate after PE
\begin{equation}\label{eq:RM}
 R_m=\tilde{R}(\mu,\tau_m,\Xi_m).
\end{equation}

\subsection{Amplifying channel}
Here, Bob detects noise described by the variable
\begin{align}
z=\sqrt{\tau} x_s \pm \sqrt{\tau-1} x_o + x_\Xi~\text{with}~\sigma_z^2=2\tau+\Xi-1
\end{align}
where $\Xi:=\tau\xi$, $\xi=\frac{(\tau-1)(\omega-1)}{\tau}$ resulting in estimator variances
\begin{align}\label{eq:sigmaXihom_amp}
\sigma_\tau^2=\frac{4\tau^2}{m}\left( 2+\sigma^2_z/(\tau V_A)\right),~~\sigma_\Xi^2=2\frac{\sigma_z^4}{m}+4\sigma_\tau^2
\end{align}
for the homodyne protocol and
\begin{align}\label{eq:sigmaXihet_amp}
\sigma_\tau^2=\frac{2\tau^2}{m}\left( 2+(\sigma_z^2+1)/(\tau V_A)\right),~~\sigma_\Xi^2=\frac{(\sigma_z^2+1)^2}{m}+2\sigma_\tau^2
\end{align}
for the heterodyne protocol. Finally, one calculates the corresponding secret key rate $R_m$ after PE  as  in~(\ref{eq:RM}).
\subsection{Classical-noise channel}
For the classical-noise channel we adopt the same analysis as in Sec.~\ref{sec:PEAtt} in addition to the assumption of
\begin{align}
\Xi&=\tau\xi=(1-\tau)\omega-(1-\tau)~~\text{with}~\lim_{\tau \rightarrow  1}\Xi=\theta.
\end{align}
This leads to the following relations for the noise variance
\begin{align}\label{eq:signalnoiseClas}
\sigma^2_z=\theta+1,
\end{align}
Therefore we obtain the worst case estimator
\begin{align}
\Xi_m=&\Xi+w\sigma_\Xi~~\text{with}~\sigma_\Xi^2=2 \frac{\sigma_z^4}{m}
\end{align}
for the homodyne protocol and
$\sigma_\Xi^2=\frac{(\sigma_z^2+1)^2}{m}$
for the heterodyne protocol. Then one obtains the corresponding secret key rate $R_m$  after PE  as in~(\ref{eq:RM}).

\section{\label{sec:composable}Composable key rates}


According to Ref.~\cite{free space}, the composable key rate takes the form
\begin{equation}
R\geq r\left[R_{m}-n^{-1/2}\Delta_{\text{AEP}}+n^{-1} \Theta \right],\label{sckeee}%
\end{equation}
where
\begin{align}
\Theta:=\left\{ \log_{2}[p\left(  1-\epsilon^2_{\text{s}}/3\right)
]+2\log_{2}\sqrt{2}\epsilon_{\text{h}}\right\},~~r=\frac{n p_\text{EC}}{N}
\end{align}
and
\begin{align}
\Delta_{\text{AEP}}&:=4\log_{2}\left(  2\sqrt
{d}+1\right)  \sqrt{\log(18/(p^2\epsilon_{\text{s}}^{4}))}, \label{AEPd}%
\end{align}
is the correction term for using the von Neumann entropy in the calculation of a finite-size rate and is dependent on the number of bins $d$ used during the discretization step of the variables.
The frame error rate $1-p_\text{EC}$ is the number of blocks with initial size $N$ that passed through the error correction (EC) step while $n=N-m$ is the portion  of signals devoted to secret key creation. With $\epsilon_\text{s}$, $\epsilon_\text{h}$, $\epsilon_\text{PE}$, and $\epsilon_\text{cor}$ we denote the smoothing parameter, the privacy amplification (hashing) parameter, the channel estimation parameter, and the EC parameter. Note that $p_\text{EC}$ is a function of $\epsilon_{EC}$ but their relation can only be evident in a specific practical implementation of the protocol. Each $\epsilon$ parameter quantifies a distance from an ideal implementation of each step of the protocol. An overall security parameter can then be calculated by composing these parameters into a sum $\epsilon=\epsilon_{\text{S}}+\epsilon_\text{cor}+\epsilon_{\text{h}}+2 p_\text{EC}\epsilon_{\text{PE}}$.  

In Figs.~\ref{fig:Att_asy},~\ref{fig:Amp_asy}, and~\ref{fig:Clas_hom}, we present results regarding the secret key rate in the composable framework for the attenuation, amplifying, and additive classical thermal noise channel, respectively. We assume conservative values for the parameters $N=10^6$, $\beta=0.9$, and $p_\text{EC}=0.8$ due to limitations that may occur in the data post-processing procedure~\cite{QKD_SIM}. However, still, the protocols provide the parties with positive rates at metropolitan distances, e.g., $\approx 10$ km [see Fig.~\ref{fig:Att_asy} (black thick dashed line)]. The security parameters have been set to $\epsilon_\text{PE}\approx 10^{-10}$ and $\epsilon_\text{s}=\epsilon_\text{h}=10^{-20}$. In addition, we chose $d=2^5$ and we optimized over $r$ and $V_A$. 

\section{Conclusion}
In this work we expanded the security of CV-QKD to all canonical forms. We studied first the asymptotic security, then  we focused on the finite-size and composable security. We first provided a compact description of the asymptotic secret-key rates of practical channels like the attenuation, amplification, and the classical-noise channels. Then our analysis discussed in more detail the impact of parameter estimation and that of other finite-size effects on the secret-key rates achievable over these channels. 
We also computed the secret-key rate for more exotic Gaussian channels finding that we either obtain an always positive key rate (for $B_1$ assuming large Gaussian modulation) or no asymptotic secret key rate (for the forms $D$ and $A_2$). This analysis can be expanded, in future works, to protocols that use squeezed and/or thermal states, protocols with discrete alphabets, or CV measurement device independent schemes, in each case by assuming links described by the previous channel classes.

\section*{Acknowledgments}
This work has been funded by the European Union’s Horizon 2020 research
and innovation program under grant agreement No 820466 (Quantum-Flagship Project CiViQ: “Continuous
Variable Quantum Communications”) and the EPSRC via the Quantum
Communications Hub (Grant No. EP/T001011/1).

\appendix
\section{$B_1$ class\label{app:b1}}
The asymptotic secret key rates for 
 the canonical form $B_1$ associated with the symplectic transformation
\begin{equation}
\mathcal{M}_{B_1}=\begin{pmatrix}
\mathbf{I} &\frac{\mathbf{I} +\mathbf{Z}}{2}\\
\frac{\mathbf{I} -\mathbf{Z}}{2} &-\mathbf{I}
\end{pmatrix}
\end{equation}
are given by
\begin{align}
&R_{\text{hom}}^\blacktriangleright(\mu)=
\frac{1}{2}\log_2\frac{\sqrt{2\mu}}{\mathrm{e}}+\frac{1}{2}h(\sqrt{2}),\label{eq:B1DDRhomQ}\\
&R_{\text{hom}}^\blacktriangleleft(\mu)
=\frac{1}{2}\log_2\frac{\sqrt{2\mu}}{\mathrm{e}},
 \\
&R_{\text{het}}^\blacktriangleright(\mu)=R_{\text{het}}^\blacktriangleleft(\mu)=\log_2\frac{\sqrt{2\mu}}{\mathrm{e}\sqrt{3}}+h(\sqrt{2}).\label{eq:B1DDRhet}
\end{align}
\end{document}